# Linking Common Vulnerabilities and Exposures to the MITRE ATT&CK Framework: A Self-Distillation Approach

Benjamin Ampel
Department of Management Information Systems
University of Arizona
Tucson, AZ, USA
bampel@arizona.edu

Sagar Samtani
Department of Operations and Decision Technologies
Indiana University
Bloomington, IN, USA
ssamtani@iu.edu

Steven Ullman
Department of Management Information Systems
University of Arizona
Tucson, AZ, USA
stevenullman@email.arizona.edu

Hsinchun Chen
Department of Management Information Systems
University of Arizona
Tucson, AZ, USA
hsinchun@arizona.edu

## ABSTRACT

Due to the ever-increasing threat of cyber-attacks to critical cyber infrastructure, organizations are focusing on building their cybersecurity knowledge base. A salient list of cybersecurity knowledge is the Common Vulnerabilities and Exposures (CVE) list, which details vulnerabilities found in a wide range of software and hardware. However, these vulnerabilities often do not have a mitigation strategy to prevent an attacker from exploiting them. A well-known cybersecurity risk management framework, MITRE ATT&CK, offers mitigation techniques for many malicious tactics. Despite the tremendous benefits that both CVEs and the ATT&CK framework can provide for key cybersecurity stakeholders (e.g., analysts, educators, and managers), the two entities are currently separate. We propose a model, named the CVE Transformer (CVET), to label CVEs with one of ten MITRE ATT&CK tactics. The CVET model contains a fine-tuning and self-knowledge distillation design applied to the state-of-the-art pre-trained language model RoBERTa. Empirical results on a gold-standard dataset suggest that our proposed novelties can increase model performance in F1-score. The results of this research can allow cybersecurity stakeholders to add preliminary MITRE ATT&CK information to their collected CVEs.

## 1 INTRODUCTION

Harmful cyber-attacks on critical cyber-infrastructure (e.g., large servers hosting confidential data) have cost on average $7.91 million per breach, leading to over 446,000,000 exposed records containing sensitive information in 2019 [22]. Thus, it is imperative to build our cybersecurity knowledge base to combat new and evolving cyber-threats. An essential component of the cybersecurity knowledge base is the Common Vulnerability and Exposures (CVE) list, overseen by the MITRE Corporation. Cybersecurity professionals commonly use CVEs to coordinate efforts to address the vulnerability. A new CVE is created whenever a security flaw is discovered in software and hardware and reported to MITRE. An example of a recent CVE is shown in Figure 1.

| CVE-ID | |
|---|---|
| CVE-2021-30502 | Learn more at National Vulnerability Database (NVD)<br>• CVSS Severity Rating • Fix Information • Vulnerable Software Versions • SCAP Mappings • CPE Information |
| **Description** | |
| The unofficial vscode-ghc-simple (aka Simple Glasgow Haskell Compiler) extension before 0.2.3 for Visual Studio Code allows remote code execution via a crafted workspace configuration with replCommand. | |
| **References** | |
| **Note:** References are provided for the convenience of the reader to help distinguish between vulnerabilities. The list is not intended to be complete. | |
| • CONFIRM:https://github.com/dramforever/vscode-ghc-simple/blob/master/CHANGELOG.md#v023<br>• CONFIRM:https://github.com/dramforever/vscode-ghc-simple/commit/bc7f6f0b857dade46ea51496d8bd1a4edef39b46<br>• MISC:https://github.com/dramforever/vscode-ghc-simple/releases<br>• MISC:https://vuln.ryotak.me/advisories/38 | |
| **Assigning CNA** | |
| MITRE Corporation | |

**Figure 1: Example CVE from cve.mitre.org**

Each CVE includes metadata such as a unique ID, a rich text description, references, and the assigning CNA. Despite their massive value to the cybersecurity community, CVEs often provide little information on how to combat the vulnerability once it is discovered. Connecting a cybersecurity risk management framework (CRMF) with tangible mitigation strategies and additional context to the CVE list to provide mitigation strategies could provide tremendous value to cybersecurity analysts and researchers.

In 2018, MITRE created the ATT&CK Matrix for Enterprise CRMF that models the tactics, techniques, and procedures (TTP) that an attacker would take when attempting to breach cyberinfrastructure [21]. The MITRE ATT&CK framework can help reliably predict and mitigate the tactics, techniques, and procedures (TTP) chain that an attacker follows after an initial breach [1]. Example tactics include "initial access," "defense evasion," and "exfiltration." Each tactic in MITRE ATT&CK comes with a mitigation strategy (e.g., user training, account management, password policies, etc.) to assist cybersecurity analysts in protecting critical cyber-infrastructure, making it an excellent CRMF for our CVE labeling task. Labeling CVEs with ATT&CK tactics requires a computational model that can analyze the textual metadata available in CVE descriptions.

Despite the tremendous benefits that both CVEs and the ATT&CK framework can provide for key cybersecurity stakeholders (e.g., analysts, researchers, educators, and managers), the two entities are currently separate. With over 158,000 CVEs existing as of the beginning of 2021, it would be a non-trivial task to manually link each one to the ATT&CK framework to gather mitigation strategies for every existing CVE.

In this study, we developed a novel framework that leverages the textual features in CVEs currently linked to an ATT&CK tactic by prior research [11] to create an ATT&CK tactic label for CVEs outside our gold-standard CVE dataset. To achieve this goal, we drew upon state-of-the-art methodologies in deep learning-based text classification literature to guide the development of a novel cybersecurity model, the CVE Transformer (CVET) model. To ensure the value of our proposed approach, we rigorously evaluated CVET against benchmark models found in related CVE data mining and cybersecurity analytics literature.

The rest of this paper is as follows. First, we review related literature to CVE data machine learning, transformers for text classification, and self-knowledge distillation. Second, we identify gaps from our literature review and pose our research question for study. Third, we detail our proposed method for labeling CVEs with MITRE ATT&CK tactics. Fourth, we summarize the empirical results of our work and discussed their implications. Finally, we conclude the work with important takeaways from the paper.

## 2 LITERATURE REVIEW

### 2.1 CVE Data Machine Learning

Large undertakings have been taken recently to use CVEs to improve other cybersecurity information systems through the use of classical machine and deep learning architectures. Authors have used the convolutional neural network (CNN) to predict CVE vulnerability severity [10] and build knowledge graphs with the CVE list, the Common Weakness Enumeration (CWE) list, and the Common Attack Pattern Enumeration and Classification (CAPEC) list [29]. However, CNNs often struggle to capture long-term dependencies in textual passages [27]. Extant literature then applied the more powerful bidirectional long short-term memory (BiLSTM) model with a self-attention mechanism to predict vulnerability severity [6] and vulnerability type (e.g., boundary condition error) [7] with higher accuracy than the CNN. More recently, researchers have leveraged the pre-trained transformer model known as BERT [23] to extract information from the prominent vulnerability database ExploitDB to enhance the textual descriptions for new CVEs. Building a model that can effectively label CVEs with ATT&CK tactics using textual descriptions requires an algorithm that can effectively represent the long text sequences found in CVE descriptions. The transformer model (and its extensions) is currently the state-of-art within text classification literature and has proven to be robust against adversarial attacks [14]. We review the transformer model in-depth to gain a deeper understanding of how it can assist in our target task.

### 2.2 Transformers for Text Classification

Introduced in 2017, the transformer model replaces the recurrent cells found in prominent text classification deep learning models (e.g., BiLSTM, LSTM) with multi-head attention mechanisms [26]. While the original design incorporates an encoder-decoder structure (for machine translation tasks), multi-class text classification only requires the encoder stack. The encoder transformer model creates an embedding from the input, passes the embedding into the transformer block, and outputs a softmax probability score of outputs. The embedding layer is a one-hot encoding with positional encodings. The transformer block consists of a multi-head attention mechanism and feed-forward layers, which has shown to greatly improve accuracy, precision, recall, and F1-score over recurrent models in benchmark tasks [8]. Recently, transformers have been used for massive pre-training language models (PTLMs) that produce state-of-the-art results in various text classification tasks [20]. These models (e.g., BERT, GPT-2) are often trained on millions of data points and contain hundreds of millions of trainable parameters. While most researchers do not own the hardware or data to create their PTLM, these models can be carefully fine-tuned [4] and distilled [12] for improved performance in specialized tasks. Knowledge distillation is an emerging paradigm that can extract key information from the parameters of a PTLM to supplement the training of a targeted model [12].

### 2.3 Self-Knowledge Distillation

Knowledge distillation (KD) combines relational knowledge from a large, pre-trained model (teacher) and a prior untrained model (student) [30]. As a result, the trained student model is often more generalizable to unseen data than a model without knowledge distillation. This design allows researchers that do not have access to the computing power required to make a massive PTLM to create highly targeted state-of-the-art models. One increasingly popular form of knowledge distillation is self-knowledge distillation (self-KD). Self-KD uses the same architecture for both the student and teacher, where knowledge transfer occurs within the same model [24]. This form of distillation creates a new model that often outperforms the original teacher model without requiring new data [32], due to distilling latent features from deeper to shallow sections of the network [33], improved feature importance weighting

[5], or modified regularization [17]. Generally, self-KD for a natural language processing task with available target labels (e.g., a text classification task) uses the weighted sum of cross-entropy (CE) loss with the correct labels and CE loss with the soft target [9].

# 3 RESEARCH GAPS AND QUESTIONS

From the extant literature, we identify several research gaps: First, many tasks have been undertaken to link CVEs to vulnerabilities, CWEs, and CAPEC, but not directly to the MITRE ATT&CK framework. The closest work to attempt this task is the BRON model [11], which does not link new CVEs, but uses existing databases to create a more holistic knowledge graph. Second, the deep learning models implemented in recent literature (e.g., CNN, BiLSTM) struggle to capture long-term dependencies in text, such as the lengthy descriptions within CVE listings. These two gaps motivate our research question:

- How can we create a novel and accurate link between CVEs and ATT&CK tactics through their textual descriptions and long-term dependencies?

# 4 PROPOSED METHOD

Our proposed methodology is comprised of three major components: (1) Data Collection and Pre-Processing, (2) the CVE Transformer (CVET) Architecture, and (3) Benchmark Experiments. Each component is further detailed in the subsequent sections.

## 4.1 Data Collection and Pre-Processing

| ATT&CK Tactic | Count of CVEs |
|---|---|
| Defense Evasion | 8,482 |
| Discovery | 6,647 |
| Privilege Escalation | 5,779 |
| Collection | 1,748 |
| Lateral Movement | 715 |
| Impact | 594 |
| Credential Access | 427 |
| Initial Access | 309 |
| Exfiltration | 137 |
| Execution | 25 |
| **Total** | **24,863** |

**Table 1: Gold-Standard Dataset CVE Distribution**

For our research, we use the dataset provided by the BRON knowledge graph [11]. As discussed earlier, there are currently more than 158,000 CVEs, and our gold-standard dataset only captures a fraction of them, making this linking task critical. The dataset leverages existing knowledge bases to link 24,863 CVEs into 10 of the 14 ATT&CK tactics. Many ATT&CK tactics do not require a vulnerability (e.g., "Resource Development" and "Command and Control"). Thus, we cannot link CVEs to them. Table 1 provides a distribution of how many CVEs are in each ATT&CK tactic category. About 91% of the data is within four tactic categories: defense evasion (8,452), discovery (6,647), privilege escalation (5,779), and collection (1,748).

To pre-process the CVE textual description, stop words were removed, non-alphanumeric characters were stripped. The remaining text was lower-cased, lemmatized, and padded to ensure proper lengths for all inputs. This sequence of pre-processing steps is common in deep learning-based text classification literature [15]. We used the pre-made RoBERTa tokenizer [16] to encode the data as input for our CVET model. Other metadata available in CVEs is not used, as it did not provide a benefit to model performance in preliminary testing.

## 4.2 CVET Architecture

*4.2.1 Model Selection.* We adapt a PTLM called RoBERTa [16] (a BERT-based model trained on longer sequences) for CVET due to the generalizability it has shown in text classification tasks [2]. While there are dozens of PTLMs to choose from for our target task, we chose RoBERTa due to the high performance it achieves while also allowing custom fine-tuning and self-KD designs.

*4.2.3 Fine-Tuning.* The standard fine-tuning process of RoBERTa and related PTLMs (e.g., BERT, GPT-2) is often unstable, with different performances depending on the random seed and dataset size [4]. Through a series of experiments, Mosbach et al. [18] identified that using the Adam optimization algorithm with bias correction led to a stable training process with improved performance compared to baseline fine-tuning (i.e., Adam without bias correction, see [3]). Therefore, we adopt this bias correction design for the CVET fine-tuning process, where bias correction is defined as:

$$\alpha_t \leftarrow \alpha \cdot \sqrt{1-\beta_2^t}/(1-\beta_1^t), \quad (2$$
$$\theta_t \leftarrow \theta_{t-1} - \alpha_t \cdot m_t/(\sqrt{v_t} + \epsilon$$

In the equation 1 and 2, $\alpha$ is the step size, $m_t$ is the first-moment estimate and $v_t$ is the second-moment estimate. From equation 1, we see that the goal of bias correction is to reduce $\alpha$ by the factor $\sqrt{1-\beta_2^t}/(1-\beta_1^t)$, which increases to 1 as $t$ increases.

*4.2.2 Self-Knowledge Distillation.* During fine-tuning of the CVET model, we use CVET as both a teacher and student. The student model (denoted as $CVET_S$) is CVET at fine-tuning time step $t$ and the teacher model (denoted as $CVET_T$) is CVET at

| Model Type | Model | Accuracy | Precision | Recall | F1-score |
|---|---|---|---|---|---|
| Classical Machine Learning | Random Forest | 63.70% *** | 35.83% *** | 37.67% *** | 36.70% *** |
| | SVM | 65.70% *** | 51.23% *** | 46.34% *** | 48.66% *** |
| | Naive Bayes | 67.30% *** | 44.22% *** | 33.92% *** | 38.16% *** |
| | Logistic Regression | 67.10% *** | 41.65% *** | 34.12% *** | 37.38% *** |
| Deep Learning | RNN | 68.45% *** | 69.66% *** | 67.30% *** | 68.46% *** |
| | GRU | 70.90% *** | 72.55% *** | 69.19% *** | 70.83% *** |
| | LSTM | 72.75% *** | 74.14% *** | **71.89%** | 72.00% *** |
| | BiLSTM | 72.55% *** | 73.71% *** | 71.71% | 72.70% *** |
| | BiLSTM with Attention | 71.41% *** | 72.52% *** | 70.32% * | 71.40% *** |
| | Transformer | 72.45% *** | 74.49% *** | 70.82% * | 73.61% *** |
| Pre-Trained Language Models | GPT-2 | 70.21% *** | 77.12% *** | 64.56% *** | 70.27% *** |
| | XLNet | 74.12% *** | 80.12% * | 68.56% *** | 73.88% *** |
| | BERT | 73.93% ** | 79.86% * | 69.41% * | 74.26% * |
| | RoBERTa | 74.42% * | **81.88%** | 68.49% ** | 74.57% * |
| Self-Distillation | **CVET** | **76.93%** | 81.38% | 71.49% | **76.18%** |

Table 2: Comparing CVET Against Benchmark Classical Machine Learning, Deep Learning, and Pre-Trained Language Models (*: p<0.05, **: p<0.01, ***: p<0.001)

fine-tuning time step $t-1$. We distill learned knowledge from $CVET_T$ to $CVET_S$ with equation 3:

$$\mathcal{L}_\theta(x,y) = CE(CVET_S(x,\theta),y) + \lambda MSE(CVET_S(x,\theta), CVET_T(x,\theta)),$$

where $x$ is the input, $y$ is the output, $\theta$ is the model's parameters, $\lambda$ is the distillation weight, CE is the cross-entropy loss, and MSE is the mean squared error loss. Simply, the self-distillation weight balances the importance of both loss functions (CE and MSE) to update the trainable parameters of $CVET_S$ based on the loss functions of $CVET_T$ and $CVET_S$, which in turn improves $CVET_T$ at the following time step. Distilling BERT-based models like this has shown improvement in many benchmark natural language processing (NLP) tasks [31].

## 5 EMPIRICAL RESULTS

The results of the experiment are shown in Table 2, and further discussed below.

### 5.1 Benchmark Experiments

We compared the proposed CVET against prevailing and state-of-the-art classical machine learning, deep learning, and pre-trained language models. Each benchmark model is commonly used for CVE data machine learning and/or text classification tasks. The models selected in each category are:

- **Classical Machine Learning**: Random Forest, SVM, Naïve Bayes, Logistic Regression
- **Deep Learning**: RNN, GRU, LSTM, BiLSTM, BiLSTM with attention, Transformer
- **Pre-Trained Language Models**: GPT-2, XLNet, BERT, RoBERTa

All classical machine learning models were implemented using the Python library scikit-learn. All deep learning models were implemented with the Python library Keras. The pre-trained language models were implemented using the Huggingface Transformers library. The CVET model used RoBERTa-large from the Huggingface library [28], and our self-distillation and fine-tuning designs were implemented in PyTorch 1.4 [19].

All models are run with 10-fold cross-validation (3 benchmark models are evaluated with the accuracy, precision, recall, and F1-score metrics, which is the standard for multi-class text classification tasks [25]. Paired t-tests are used to identify if statistically significant differences exist between CVET and each benchmark method. Due to our unbalanced dataset, the discussion focuses on the F1-score (which is more resilient against skewed distributions than the other metrics) [13].

### 5.2 Results and Discussion

From Table 2, we make four key observations about the results of our experiments. First, the four classical machine learning models had the lowest F1-scores, ranging from 36.70% for Random Forest to 48.66% for SVM. Second, all deep learning models reached a higher F1-score than the classical machine learning models. The transformer model had the highest F1-score (73.61%) among the deep learning models. The transformer does not have an internal recurrent mechanism like the other models, which suggests its multi-head attention architecture helped improve the text classification performance. Third, all of the PTLMs marginally improved upon the transformer model in F1-score. Baseline RoBERTa improved over the transformer by 0.96% (from 73.61% to 74.57%). These results suggest that pre-training the transformer can improve the performances of specialized text classification tasks. Finally, our proposed CVET model outperformed all other PTLMs, deep learning models, and classical machine learning models in F1-score (76.18%). The differences were significant at $p<0.05$ or less in all models. Our CVET model also achieved the best accuracy (79.93%) versus all other models. These results suggest that our fine-tuning and self-KD design assisted in performance improvement in the target task.

## 6 CONCLUSION

In this study, we developed a novel self-distillation approach to automatically label CVEs with their associated ATT&CK tactic. The CVET model fine-tuning process included an Adam loss function with added bias correction and a self-KD design. Our model was evaluated with a series of experiments against state-of-the-art models in classical machine learning, deep learning, and pre-trained language models. Results indicated that the CVET model offers a significant benefit to labeling CVEs with MITRE ATT&CK tactics over baseline non-distillation techniques. Our architecture can greatly assist the cybersecurity community by creating an immediate link between a heavily utilized cybersecurity risk management framework and critical vulnerabilities. This can be implemented into key cybersecurity stakeholders' workflow to associate vulnerabilities found in their scanners to the MITRE ATT&CK framework for additional information on how to combat the vulnerability.

We identify two potential future directions for related work in this domain. First, we would like to expand the connection of CVEs to other prominent CRMFs. Potential options include the CAPEC list and the National Institute of Standards and Technology (NIST) framework. Such connections can broaden the mitigation strategies provided by this work when a CVE is discovered. Second, we plan to look into other more refined textual data representation techniques (e.g., novel word embedding strategies, synonym/homonym generation, POS tagging) to improve the features of our textual inputs.